\documentstyle[12pt]{article}
\def\headline={\ifnum\pageno=1\firstheadline\else
\ifodd\pageno\rightheadline \else\leftheadline\fi\fi}
\def\firstheadline{\hfil}
\def\rightheadline{\hfil}
\def\leftheadline{\hfil}
\def\footline={\ifnum\pageno=1\firstfootline\else\otherfootline\fi}
\def\firstfootline{\rm\hss\folio\hss}
\def\otherfootline{\hfil}

\font\twelvebf=cmbx10 scaled\magstep 1
\font\twelverm=cmr10 scaled\magstep 1
 1

\font\tenrm=cmr10
\font\tenit=cmti10

\input{epsf}

\begin{document}

\centerline{\twelvebf ISING SPINS ON THE LABYRINTH}
\vglue 1cm
\centerline{\twelverm UWE GRIMM,}
\baselineskip=13pt
\centerline{\tenit Instituut voor Theoretische Fysica,
Universiteit van Amsterdam,}
\baselineskip=12pt
\centerline{\tenit Valckenierstraat 65, 1018 XE Amsterdam, The Netherlands}
\baselineskip=13pt
\vglue 0.6cm
\centerline{\twelverm MICHAEL BAAKE \ and \ HARALD SIMON}
\baselineskip=13pt
\centerline{\tenit Institut f\"ur Theoretische Physik,
Universit\"at T\"ubingen}
\baselineskip=12pt
\centerline{\tenit Auf der Morgenstelle 14, 72076 T\"ubingen, Germany}
\vglue 0.8cm
\centerline{\tenrm ABSTRACT}
\baselineskip=13pt
\centerline{\parbox[t]{33.2pc}{\tenrm\baselineskip=12pt
We consider a zero-field Ising model 
defined on a quasiperiodic graph, the
so-called Labyrinth tiling. Exact information about the critical behaviour
is obtained from duality arguments and the subclass of models which yield 
commuting transfer matrices. For the latter, the magnetization is independent 
of the position and the phase transition between ordered and disordered phase 
belongs to the Onsager universality class. In order to obtain information about 
the generic case, we calculate the magnetization for a series of couplings by 
standard Monte-Carlo methods.}}
\vspace{0.2truein}
\baselineskip=14pt
    
\subsubsection*{1. Introduction}

\mbox{\indent}Not too much is known about the critical behaviour of statistical
systems on non-periodic graphs. Although certain cases may still be treated
by commuting transfer matrices (see Ref.~\ref{BGB} and references therein)
using so-called $Z$-invariance arguments$^{\ref{Baxter},\ref{Baxter86}}$,
these may well not be representative. For instance, for the case at hand,
the local magnetization turns out to be position-independent, what one 
certainly does not expect to happen for a general (ev.\ random) 
distribution of the coupling constants.

The example considered in this note, the Ising model on the so-called
Labyrinth tiling, has recently been investigated in detail$^{\ref{BGB}}$
by means of duality arguments and commuting transfer matrices. Here, we  
briefly review the relevant results and supplement these by 
first numerical investigations.

\subsubsection*{2. The Model}

\mbox{\indent}The silver mean chain is obtained by repeated application of the
two-letter substitution rule $(a\rightarrow b, b\rightarrow bab)$
to the letter $a$. From this, the Labyrinth tiling$^{\ref{SireMossSad89}}$  
can be constructed by considering an orthogonal Cartesian product of two 
identical silver mean chains in the proper geometric 
representation$^{\ref{BGB},\ref{SireMossSad89}}$ and connecting points
on one of the two subgrids of the resulting rectangular grid, see Fig.~1.

In this way, we obtain a tiling which consists of three different tiles.
As a graph, it has the topology of the square lattice, but contains edges
of three types (or eight if one accounts for orientations). 
It is therefore natural to assign individual coupling constants to 
different types of edges, defining in this way a nearest-neighbour coupling
for the Ising spins $\sigma_{i,j}\in\{\pm 1\}$ which we place 
on the vertices of the graph. We denote the ferromagnetic couplings 
(in units of $k_{B}T$) by $K_{xy}^{}$ and $L_{xy}^{}$, where 
$xy\in\{aa,ab,ba,bb\}$ labels the abscissa and ordinate of the
corresponding rectangle in the underlying grid and the letters $K$ and
$L$ refer to the two different diagonals$^{\ref{BGB}}$.

The proper periodic approximants for finite systems are constructed as above,
but from a periodic grid. This is defined by identifying the first and
last letter of a word obtained by applying the substitution rule a certain
number of times. This ensures that no additional tiles or 
vertex configurations are created$^{\ref{BGB}}$.

\subsubsection*{3. Exact Results}

\mbox{\indent}Let us restrict to the case of non-vanishing ferromagnetic 
couplings that are
uniformly bounded from above and below by finite constants. Under these
assumptions, the Peierls argument$^{\ref{Peierls},\ref{BGB}}$ guarantees
the existence of at least one phase transition.

Some more information about the critical behaviour can be
obtained from a duality argument. Assuming that there is only
a single transition, it must occur on the self-dual surface in the
space of coupling constants, which is given by$^{\ref{Baxter86},\ref{BGB}}$
\begin{equation} \label{selfdual}
  S_{xy}\; := \;\sinh(2 K_{xy}^{\mbox{}})\, \sinh(2 L_{xy}^{\mbox{}}) \; =\; 1
\end{equation}
for all index pairs $xy\in\{aa,ab,ba,bb\}$. 

Even more, the model is exactly solvable in the sense of commuting transfer 
matrices$^{\ref{Baxter}}$ in a subspace defined by the four equations 
$S_{xy}=\Omega$ (for the possible index pairs $xy\in\{aa,ab,ba,bb\}$)
plus one additional equation, see eq.~(4.5) in Ref.~\ref{BGB}. 
The corresponding coupling constants can be parametrized explicitly
in terms of elliptic functions. For a given bond, the argument is the
difference of two rapidity parameters$^{\ref{Baxter78},\ref{Baxter86}}$ 
that are attached to the two lines which intersect on the bond, see Fig.~1.  
\begin{figure}[hb]
\centerline{\epsfysize=135mm \epsfbox[100 170 530 590]{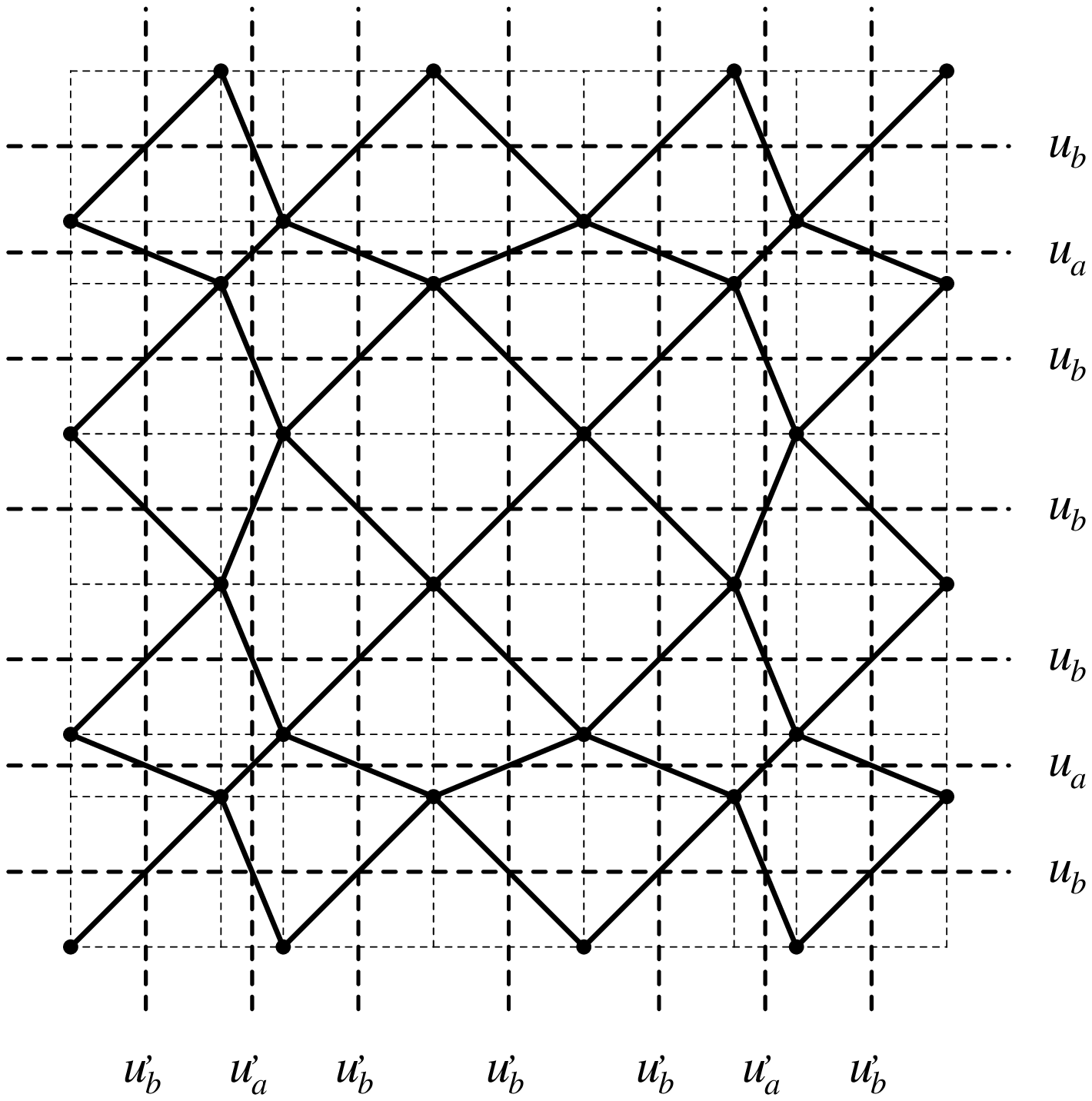}}
\centerline{\tenrm Fig.~1. A finite patch of the Labyrinth with underlying 
            grid and rapidity lines.}
\end{figure}
In the three-dimensional solvable subspace, the model shows (in terms of the 
temperature-like variable $\Omega^2$) a single second-order phase transition 
at $\Omega=1$ (i.e., on the intersection of the solvable and the self-dual 
surface, compare Eq.~\ref{selfdual}) which belongs to the Onsager universality 
class. In particular, the local magnetization $\langle\sigma\rangle$ turns out 
to be {\em position-independent}\/ and shows, in the thermodynamic limit, the 
critical singularity$^{\ref{BGB},\ref{Baxter86}}$
\begin{equation} \label{mag}
\langle\sigma\rangle \;\; = \;\; 
\left\{ \begin{array}{l@{\hspace*{8mm}}l}
{(1-\Omega^{-2})}^{1/8} & \mbox{if $\Omega^2>1$} \\
 0 & \mbox{if $\Omega^2\leq 1$} \end{array} \right.
\end{equation}
at $\Omega=1$ governed by the magnetic exponent $\beta=1/8$ of the Ising model.
Furthermore, one can calculate the free energy by essentially counting bond
frequencies$^{\ref{BGB}}$. This is due to the ``mobility'' of the rapidity
lines$^{\ref{Baxter78}}$ which is a consequence of the Yang-Baxter equation.
Note that the periodic boundary conditions guarantee that moving rapidity lines
does not create any surface contributions in our case.

Clearly, the latter result reflects the severe restrictions imposed by 
integrability. It is an interesting question whether the converse is also true, 
i.e., whether the position-independence of the local magnetization is 
sufficient for solvability. For a generic choice of couplings, one certainly 
expects the local magnetization to depend on the neighbourhood. This poses 
the question whether the solvable case is representative at all --- a partial 
answer to which can be obtained by numerical investigation of non-integrable 
cases. 

\subsubsection*{4. Numerical Results}

\mbox{\indent}As a first approach, we consider the dependence of the 
local magnetization 
on the position of the spin while digressing from the solvable surface.
The simplest scenario occurs when going from the periodic case (i.e.,
all couplings equal) to the case of three different couplings according
to the length of the bonds, i.e., $K_{aa}=L_{aa}=J_{s}/k_{B}T$, 
$K_{ab}=K_{ba}=L_{ab}=L_{ba}=J_{m}/k_{B}T$ and $K_{bb}=L_{bb}=J_{l}/k_{B}T$, 
where the subscripts $s$, $m$ and $l$ refer to short, medium and long bonds, 
respectively. We consider the periodic approximant which is defined by the 
word of length 41 obtained after five applications of the substitution rule 
to the initial letter $a$. For this patch of 1600 sites, we estimated the 
magnetization at three different sites (where we chose representatives of the 
three different vertex configurations, see Fig.~2) 
by means of the Swendsen-Wang Monte-Carlo algorithm$^{\ref{Bin}}$. 
\begin{figure}[hb]
\centerline{\epsfysize=50mm \epsfbox[80 330 500 460]{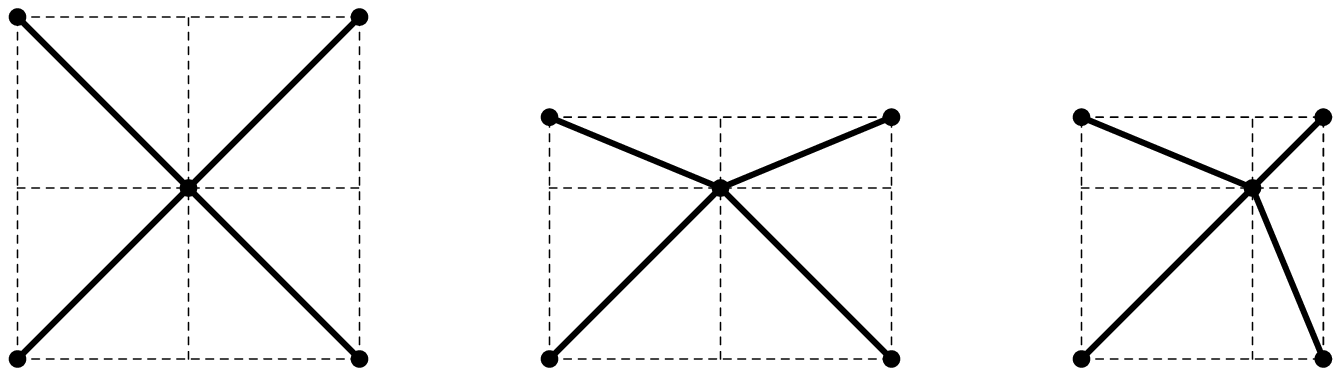}}
\centerline{\tenrm Fig.~2. The three different vertices of the Labyrinth.}
\end{figure}
The result is shown in Fig.~3, where the 
abscissa displays $T/T_c$ and the ordinate the normalized magnetization. 
Here, the three sets of coupling constants were chosen as follows.
Fig.~3(a) corresponds to $J_{s}=J_{m}=J_{l}$, in Fig.~3(b) we used
$J_{s}/J_{m}=6/5$, $J_{l}/J_{m}=4/5$ and Fig.~3(c) displays the results
for $J_{s}/J_{m}=7/5$ and $J_{l}/J_{m}=3/5$. In order to keep the critical
temperature approximately constant, we adjusted the coupling constants 
such that the average coupling (per bond) for the three cases is the same.

\begin{figure}[ht]
\centerline{\epsfysize=45mm \epsfbox[80 340 540 460]{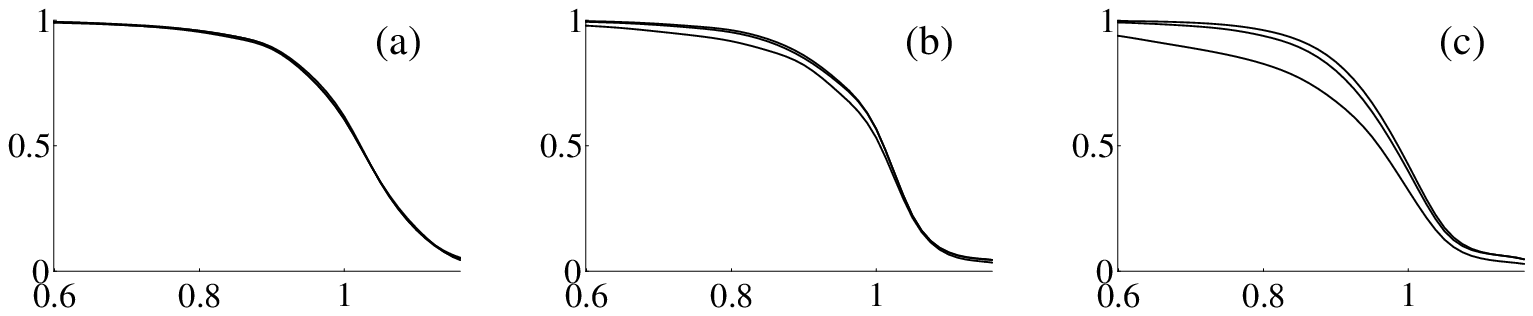}}
\centerline{\tenrm Fig.~3. Local magnetization at three positions 
                           for different couplings.}
\end{figure}

One can clearly see that the magnetization is site-independent in the periodic 
case (a), as it must be, but develops pronounced site-dependence as the 
difference of the three couplings increases (from (b) to (c)). 
On the other hand, we cannot decide whether we get more than one
point of phase transition or a critical region, although our
calculations seem to support the expectation that the critical point
is unique.

\subsubsection*{5. Concluding Remarks}

\mbox{\indent}To get a better understanding of critical phenomena 
on non-periodic
graphs with long-range order (which are in between the periodic and
the random case), we have investigated the classical Ising model on
a 2D quasiperiodic tiling. The example chosen, the so-called Labyrinth,
can be considered as a quasiperiodic modulation of the square lattice.
The Ising model is exactly solvable on a three-dimensional subspace of
the coupling space considered, which contains the periodic case.
Solvability resulted in site-independence of the local magnetization,
while the generic case shows clear dependence on the local neighbourhood.
A conclusive statement on the phase structure and the nature of critical 
behaviour requires further investigation of the model by algebraic and 
numerical means.

\subsubsection*{6. Acknowledgements}

\mbox{\indent}We thank R.J.~Baxter, B.~Nienhuis and
P.A.~Pearce for discussions.
U.G.\ is grateful for financial support from the 
Samenwerkingsverband FOM/SMC Mathematische Fysica.

\subsubsection*{7. References}

{\small
\begin{enumerate}
\baselineskip=12pt
\parskip=0.9pt

\newcommand{\bibi}{\item}

\bibi \label{BGB}
M.~Baake, U.~Grimm and R.~J.~Baxter,
``A Critical Ising Model on the Labyrinth'',
{\em Int.\ J.\ Mod.\ Phys.}\ {\bf B8} (1994) 3579--600;
reprinted in: {\em Perspectives on Solvable Models},
eds.\ U.~Grimm and M.~Baake (World Scientific, Singapore, 1994),
p.~131--52.

\bibi \label{Baxter78}
R.~J.~Baxter,
``Solvable eight-vertex model on an arbitrary planar lattice'',
{\em Philos.\ Trans.\ R.\ Soc.\ London} {\bf 289} (1978) 315--46.

\bibi \label{Baxter}
R.~J.~Baxter,
{\em Exactly Solved Models in Statistical Mechanics}\/
(Academic Press, London, 1982).

\bibi \label{Baxter86}
R.~J.~Baxter,
``Free-fermion, checkerboard and $Z$-invariant lattice models
  in statistical mechanics'',
{\em Proc.\ R.\ Soc.\ London}\ {\bf A404} (1986) 1--33.

\bibi \label{Bin}
J.~J.~Binney, N.~J.~Dowrick, A.~J.~Fisher and M.~E.~J.~Newman,
{\em The Theory of Critical Phenomena}\/ (Clarendon Press,
Oxford, 1992).

\bibi \label{Peierls}
R.~Peierls,
``On Ising's model of ferromagnetism'',
{\em Proc.\ Cambridge Philos.\ Soc.}\ {\bf 32} (1936) 477--81.

\bibi \label{SireMossSad89}
C.~Sire, R.~Mosseri and J.-F.~Sadoc,
``Geometric study of a 2D tiling related to the octagonal 
 quasiperiodic tiling'',
{\em J.\ Phys.\ (France)} {\bf 50} (1989) 3463--76.

\end{enumerate}
 }
\end{document}